\documentclass[12pt]{article}

 \newread\testifexists
 \def\GetIfExists #1 {\immediate\openin\testifexists=#1
     \ifeof\testifexists\immediate\closein\testifexists\else
     \immediate\closein\testifexists\input #1\fi}

 \usepackage{gthstyle}
 \usepackage{amsfonts}
 \usepackage{amssymb}
 \immediate\openin\testifexists=e
     \ifeof\testifexists\immediate\closein\testifexists\else
     \immediate\closein\testifexists\input e\fipsf

 \def\Bbb#1{\setbox0=\hbox{$\tt #1$}  \copy0\kern-\wd0\kern .1em\copy0}

 \def\bbf#1{\setbox0=\hbox{$#1$} \kern-.025em\copy0\kern-\wd0
         \kern.05em\copy0\kern-\wd0 \kern-.025em\raise.0433em\box0}

 \immediate\openin\testifexists=a
     \ifeof\testifexists\immediate\closein\testifexists\else
     \immediate\closein\testifexists\input a\fimssym.def  

 \def\a{\alpha}      \def\b{\beta}   \def\g{\gamma}      \def\G{\Gamma}
 \def\d{\delta}      \def\D{\Delta}  \def\e{\varepsilon} \def\et{\eta}
 \def\k{\kappa}      \def\l{\lambda} \def\L{\Lambda}     \def\m{\mu}
 \def\f{\phi}            \def\vv{\varphi}    \def\n{\nu}
 \def\j{\psi}            \def\r{\varrho}     \def\s{\sigma}

 \def\w{\omega}      \def\W{\Omega}  

    \def\LL{{\mathcal L}} \def\OO{{\mathcal O}} \def\DD{{\mathcal D}}\def\NN{{\mathcal N}}
 \def\pa{\partial} \def\ra{\rightarrow}
  
 \def\dd{{\rm d}}     

 \def\qu{\ {\buildrel {\displaystyle ?} \over =}\ }
 \def\deff{\ {\buildrel{\rm def}\over{=}}\ }
 \def\iss{\ =\ }

 \def\fract#1#2{{\textstyle{#1\over#2}}}
 \def\ffract#1#2{\raise .2 em\hbox{$\scriptstyle#1$}\kern-.3em/
                 \kern-.2em\lower .15 em \hbox{$\scriptstyle#2$}}
 
 \def\half{\fract12} \def\quart{\fract14}

 \newcommand{\tl}[1]{\tilde{#1}}              \newcommand{\Tr}{{\mbox{Tr}}\,}
                     \newcommand{\fn}{\footnote}
 \newcommand{\nn}{\nonumber\\[2pt]}             
 \newcommand{\be}{\begin{eqnarray}}             \newcommand{\ee}{\end{eqnarray}}
 \newcommand{\bi}[1]{\begin{itemize}\item[#1]}         \newcommand{\itm}[1]{\item[#1]}
       \newcommand{\ei}{\end{itemize}\noindent}
 \newcommand{\eqn}[1]{(\ref{#1})}

 \newcommand{\crlb}[1]{\label{#1}\\[2pt]}
 \newcommand{\eela}[1]{\quad\hbox{\scriptsize{#1}}\label{#1}\end{eqnarray}}
 \newcommand{\eelb}[1]{\label{#1}\end{eqnarray}}
 
 \newcommand{\newsecb}[2]{\section{#1}\label{#2}\setcounter{equation}{0}}
 
 \newcommand{\nolabels} {\def\eel{\eelb} \def\crl{\crlb} \def\newsecl{\newsecb}}

\newcommand\publishversion{\nolabels\setlength{\textheight}{9in}\setlength{\oddsidemargin}{0in}
    \setlength{\textwidth}{6.3in}\setlength{\topmargin}{-0.1in}}

\publishversion
\begin{document} \begin{titlepage}

\title{\normalsize \hfill ITP-UU-11/14 \\ \hfill SPIN-11/08\\
\vskip 20mm \Large\bf A class of elementary particle models without any adjustable real parameters}

\author{Gerard 't~Hooft}
\date{\normalsize Institute for Theoretical Physics \\
Utrecht University \\ and
\medskip \\ Spinoza Institute \\ Postbox 80.195 \\ 3508 TD Utrecht, the Netherlands \smallskip \\
e-mail: \tt g.thooft@uu.nl \\ internet: \tt http://www.phys.uu.nl/\~{}thooft/}

\maketitle

\begin{quotation} \noindent {\large\bf Abstract } \medskip

{\small{Conventional particle theories such as the Standard Model have a number of freely adjustable coupling
constants and mass parameters, depending on the symmetry algebra of the local gauge group and the
representations chosen for the spinor and scalar fields. There seems to be no physical principle to determine
these parameters as long as they stay within certain domains dictated by the renormalization group. Here
however, reasons are given to demand that, when gravity is coupled to the system, local conformal invariance
should be a spontaneously broken exact symmetry. The argument has to do with the requirement that black holes
obey a complementarity principle relating ingoing observers to outside observers, or equivalently, initial
states to final states. This condition fixes all parameters, including masses and the cosmological constant.
We suspect that only examples can be found where these are all of order one in Planck units, but the values
depend on the algebra chosen. This paper combines findings reported in two previous preprints,[1][2] and puts
these in a clearer perspective by shifting the emphasis towards the implications for particle
models.}}\end{quotation}
  \vfill \flushleft{April 22, 2011}

\end{titlepage}

\eject

\def\EH{\mathrm{\,EH}}  \def\mat{\mathrm{\,mat}} \def\tot{\mathrm{\,total}} \def\kin{\mathrm{\,kin}}
\def\mass{\mathrm{\,mass}} \def\intt{\mathrm{\,int}} \def\eff{\mathrm{\,eff}}


\newsecl{Introduction: splitting the functional integral}{intro}
The Einstein-Hilbert action of the gravitational part of a generally covariant theory reads
 \be S^{\,\mathrm{EH}}=\int\dd^4 x\,  {\textstyle\sqrt{-g}\over 16\pi G_N} (R-2\L)\ , \eel{EH}
where for later convenience we already added a possible cosmological term \(-2\L\), and \(G_N\) is
Newton's constant. The total reparametrization invariant action is written as
 \be S^{\,\mathrm{total}}=S^{\,\mathrm{EH}}+\int\dd^4x\sqrt{-g}\LL^\mat\ , \eel{Stotal}
where the matter Lagrangian \(\mathcal{L}^\mat\) is written in a generally covariant manner using
the space-time metric \(g_{\m\n}(x)\). To study conformal invariance, it is useful to split the
metric tensor \(g_{\m\n}(x)\) as follows:
 \be g_{\m\n}(x)\deff\w^2(x)\,\hat g_{\m\n}(x)\ , \eel{gsplit}
where \(\hat g_{\m\n}(x)\) may be subject to some additional constraint such as
 \be \det(\hat g)=-1\ , \eel{scalegauge}
\textit{besides} imposing a gauge condition for each of the \(n=4\) coordinates. Condition
\eqn{scalegauge} is not at all necessary; it is just one of the possible choices for fixing the
gauge concerning the following local gauge transformation,
 \be \hat g_{\m\n}(x)\ra \l^2(x)\,\hat g_{\m\n}(x)\ ,\qquad \w(x)\ra \l^{-1}(x)\,\w(x)\ .
 \eel{localconftrf}
This transformation, which is \emph{not} a coordinate transformation, will be referred to as a \emph{local
conformal transformation}. We note that a special way to fix the gauge here would simply be \be \w(x)\ra 1\ ,
\eel{omegaone} in which case the conventional theory is reobtained. On the other hand, however, one could
note that, if \(\l(x)\) is taken to be constant, this transformation is a scale transformation for the
modified distance unit \(\dd \hat s^2=\hat g_{\m\n}\dd x^\m\dd x^\n\). Furthermore, if \(\l(x)=\r^2/x^2\),
where \(\r\) is a fixed constant, we can transform a flat \(\hat g_{\m\n}\) back into a flat space-time
metric if it is combined with the space-time transformation
 \be x^\m\ra\r^2\,x^\m/x^2\ ;\eel{specialconftrf}
this will always be referred to as a \emph{special \emph{(or global)} conformal transformation}.

The gauge choice \eqn{omegaone} could be interpreted as the `unitarity gauge' while a conformal
Higgs mechanism takes place: the ``vacuum expectation value" of \(\w\) is one.

Let us write the functional integration procedure as
 \be\int\DD g_{\m\n}(x)\,e^{S^\mathrm{\,coordfix}}[\,\cdots]=\int\DD\w(x)\int\DD\hat g_{\m\n}(x)\,
 e^{S^\mathrm{\,conffix+coordfix}}[\,\cdots] , \eel{functintsplit}
where \(S^\mathrm{\,coordfix}\) is the gauge fixing constraint for the local coordinate reparametrization
ambiguity, including the associated Faddeev Popov ghost, and \(S^\mathrm{\,conffix+coordfix}\) fixes both the
coordinate reparametrization ambiguity and the conformal gauge ambiguity\fn{A fine choice would be, for
instance, \(\pa_\m\hat g^{\m\n}=0\) \emph{and} \(\det(\hat g_{\m\n})=-1\). Again, the usual Faddeev Popov
quantization procedure is assumed.}. The latter may or may not be chosen to depend on \(\w(x)\). The
interesting case is if it does not, thus ruling out the gauge choice \eqn{omegaone}. In that case, we can
carry out the functional integral over \(\w(x)\) to obtain an effective action in terms of \(\hat g_{\m\n}\)
and the matter fields which is still totally invariant under local conformal transformations. It is this
effective conformally invariant theory that can be used to produce a black hole theory with a complemantarity
principle.\cite{GtHcompl} As we shall see, the constraint that makes this complemantarity principle work,
also gives constraints on all matter field interactions, so that no freely adjustable constants of nature
remain.

In the standard perturbation expansion, the integration order does not matter. Also, if dimensional
regularization is employed, the choice of the functional metric in the space of all fields
\(\w(x)\) and \(\hat g_{\m\n}(x)\) is unambiguous, as any effects that depend on this choice are
cancelled out by the Faddeev Popov ghost contributions.

If there had been no divergences in the theory at all, one would have expected the following
scenario:
 \bi{-} After integrating over all scale functions \(\w(x)\), but not yet over \(\hat g_{\m\n}\),
the resulting effective action in terms of \(\hat g_{\m\n}\) and the matter fields should be
expected to stay locally conformally invariant, \textit{i.e.} if we would split \(\hat g_{\m\n}\)
again as in Eq.~\eqn{gsplit},
 \be \hat g_{\m\n}(x)\qu \hat\w^2(x) \hat{\hat g}_{\m\n}\ , \eel{gdubbelsplit}
no further dependence on \(\hat\w(x)\) should be expected.
 \itm{-} Therefore, the effective action should now describe a conformally invariant theory, both for gravity
and for matter. Because of this, the effective theory might be expected to be renormalizable, or
even finite! If any infinities do remain, one might again employ dimensional renormalization to
remove them. \ei
 However, this expectation is jeopardized by a conspicuous difficulty: the \(\w\) integration,
as well as the matter integrations and the \(\hat g_{\m\n}\) integration, are indeed ultraviolet
divergent.\cite{BDeWitt}---\cite{GtHVeltman} The fact that the original theory \eqn{EH} was not
renormalizable is here reflected in the fact that counter terms will be needed that do not occur
in the lowest order terms for \(\hat g_{\m\n}\), as we shall see. On top of this, we shall
encounter anomalies that violate local conformal invariance.\cite{capperduff}\cite{Duff} What is
the physical meaning of these anomalies? Can we, and should we, cure them by having them cancel
out? This is the question that will be addressed. The answer will lead us to new constraints on
the system: the renormalizable particle interactions will be subject to algebraic conditions.

In Section~\ref{confmatter.sec} the theory in its new jacket is being discussed. It is found that the \(\w\)
integration is technically identical to the integration over a conventional, renormalizable scalar field. We
see how local conformal invariance continues to be a formal local symmetry of the system, which might be said
to be \emph{spontaneously broken}, akin to the Higgs mechanism. If there is any difficulty in quantizing and
renormalizing gravity, it is in the integration over \(\hat g_{\m\n}\). In this formalism, somewhat
disturbingly, there is no tree diagram contribution to the \(\hat g_{\m\n}\) propagator; it all must come
from the loops. This might still have been acceptable, if there had not been any anomaly.

There are, in fact, two important types\fn{A third well-known anomaly is associated to the
Gauss-Bonnet term, a topological term in the Lagrangian that does not contribute to physical
effects in topologically trivial space-times, see Eq.~\eqn{GB} in Appendix~\ref{Weyl}. Since, even
in the case of black holes, our theory forces space and time to be topologically trivial, this
third anomaly appears to play no direct role here, and it will not further be discussed.} of
anomalies. In Sections~\ref{scalefint} and \ref{matterdiv}, the first of these is discussed: an
anomaly that occurs in a curved background of a \(\hat g_{\m\n}\) metric. We begin by strictly
adhering to the standard formalism of quantum gravity; anomalies are found that, in the standard
formalism, are taken for granted. Then, in Section~\ref{flat}, an anomaly that is already there in
a flat background is considered. We observe that, apart from the anomalies, the standard canonical
theory can be viewed upon as a conformal Higgs theory. It is here that we insist that, instead, we
should be dealing with an \emph{exact} conformal Higgs mechanism, where local conformal invariance
should \emph{not} be broken by anomalies, and this condition forces us to introduce a new physical
constraint. We shall observe that this constraint can be easily imposed for the second type of
anomalies. Thus, the required anomaly cancelation can be enforced in a flat, or at least
Ricci-flat background. This is harder to achieve for the first type of anomalies, and we conclude
that the functional integral over \(\hat g_{\m\n}\) is not yet well understood; evidently, our
theory is not anywhere close to a ``theory of everything", but in Section~\ref{beta} we do find a
result when local conformal invariance is asked for in a flat background: all parameters that
previously have been looked upon as freely adjustable real parameters -- or at least freely
adjustable within some range -- are now all fixed. The calculation is done in
Section~\ref{betanul}. This will confront us with an other problem: the \emph{hierarchy problem}:
why do dimensionless ratios of physical constants in our world take extremely large or extremely
small values? This problem, and other features, are briefly discussed in Section~\ref{disc}.
Conclusions are described in Section~\ref{concl}.

\newsecl{The locally conformal formalism including matter fields}{confmatter.sec}

In terms of the fields \(\w(x)\) and \(\hat g_{\m\n}(x)\), the Einstein-Hilbert action reads
 \be S^\EH=\int\dd^4x{1\over 2\k^2}\Big(\hat R\,\w^2+6\,\hat
 g^{\m\n}\pa_\m\w\pa_\n\w-2\L\w^4\Big)\ , \eel{EHsplit}
where \(\k^2=8\pi G_N\), and \(\hat R\) is the Ricci scalar associated to \(\hat g_{\m\n}\). When adding the
matter lagrangian, it is convenient to split that into conformally invariant kinetic parts, mass terms, and
interaction terms. In a simplified notation (later, in Section~\ref{flat}, we will be more precise), one has
 \be \f^\mat&=&\{A_\m(x), \j(x),\bar\j(x),\vv(x)\}\ ,\qquad\LL^\mat=\LL^\kin+\LL^\mass+\LL^\intt\ ;\crl{matterfields}
 \LL^\kin&=&-\quart\hat g^{\m\a}\hat g^{\n\b}F_{\m\n}F_{\a\b}-\half\hat
 g^{\m\n}\pa_\m\vv\,\pa_\n\vv-\fract1{12}\hat R\vv^2-\bar\j \g^\m\hat D_\m\j\ ;\crl{conformalL}
 \LL^\mass&=&-\half m_s^2\w^2\vv^2-\bar\j\,\w\, m_d\,\j\ . \eel{massL}
Here, \(F_{\m\n}=\pa_\m A_\n-\pa_\n A_\m\); the kinetic term for the Dirac fields is shorthand for the
corresponding expression using a vierbein field \(\hat e_\m^a\) for the metric \(\hat g_{\m\n}\) with its
associated connection field. \(m_s\) stands short for the scalar masses and \(m_d\) for the Dirac masses. The
term \(\fract1{12}\hat R\vv^2\) could be removed by a field redefinition, but is kept here for convenience,
making the scalar lagrangian explicitly conformally invariant.

Then, we have the matter interaction terms, such as quartic scalar interaction terms and Yukawa interaction
terms,
 \be \LL^\intt =-\fract 1{4!}\l \vv^4-y\,\bar\j\vv\j\ +\ \cdots, \eel{Lint}
where \(y\) stands for the Yukawa interaction constants (possibly including \(\g^5\) terms) and \(\cdots\)
for possible Yang-Mills interactions.

The relevant local conformal transformations of the matter fields and the vierbein are\fn{To understand how
spinors \(\bar\j,\ \j\) transform, one has to realize that their covariant derivative \(\hat D_\m\) depends
on the vierbein field \(\hat e^a_\m\) which is therefore not invariant under local conformal transformations.
The spinorial kinetic term in Eq.~\eqn{conformalL} can then be seen to be invariant under \eqn{conftrfmatter}
up to a total derivative.}
 \be \vv(x)&\ra&\l^{-1}(x)\,\vv(x)\ , \nn
  \hat e_\m^a(x)&\ra&\l(x)\,\,\hat e_\m^a(x)\ , \nn
 \bar\j(x)&\ra& \l^{-3/2}(x)\,\bar\j(x)\ , \nn
 \j(x)&\ra&\l^{-3/2}(x)\,\j(x)\ . \eel{conftrfmatter}

We notice now that, by construction, all contributions to the total lagrangian are invariant under these
transformations, when combined with \eqn{localconftrf}. It is no coincidence that the contribution of the
\(\w\) field, Eq.~\eqn{EHsplit}, has the same form as that of the conformal scalar fields \(\vv\), as they
both transform the same way. In particular, we see the resemblance between the cosmological term and the
\(\l\vv^4\) coupling. To exploit this, we write
 \be\w(x)= i\,\tl\k\,\eta(x)\ ,\qquad\tl\k^2=\fract 1{6}\k^2=\fract 43\pi G_N\ , \eel{omegaeta}
so that \(\eta(x)\) acts exactly as a scalar field. The factor \(i\) indicates that the functional integral
over \(\w\) has to go over a contour parallel to the imaginary axis, a well-known feature of canonical
quantum gravity, important particularly if one wishes to make the Wick rotation to a Euclidean background
spacetime.

One important distinction between the \(\eta\) field and the other scalar fields \(\vv(x)\) remains: all
those interactions that are odd in the \(\eta\) field must have purely imaginary coupling constants, a
peculiar consequence of the unitarity requirement of quantum gravity (terms such as the Dirac mass term in
Eq.~\eqn{massL} must be real).

Since \eqn{EHsplit} is the \emph{entire} the Einstein-Hilbert action, we see that no kinetic term survives
for the \(\hat g_{\m\n}\) field. One might be tempted to invent a kinetic term, but then we have to realize
that this may jeopardize unitarity; we know that without any such term, the standard interactions are unitary
perturbatively. We retun to these questions at the end of Section~\ref{matterdiv}. In Appendix \ref{Weyl} we
briefly remind the reader of the local conformal properties of the curvature fields. The Riemann and Ricci
curvatures do not transform in a simple way, but those parts of the Riemann curvature that are orthogonal to
the Ricci curvature, called the Weyl curvature, a 4-tensor with 10 independent components, does transform
into itself. In principle, this allows us to write exactly one kind of kinetic term for \(\hat g_{\m\n}\),
which would be locally conformally invariant and hence renormalizable, but unfortunately not manifestly
unitary. Possible procedures to restore unitarity for such models were suggested by
Mannheim\cite{PM}\cite{BM}, but these ar controversial; operators referring to the metric fail to be
hermitean, a serious flaw according to many researchers.

By construction, all terms \eqn{EHsplit}---\eqn{Lint} of our Lagrangian are invariant under the local
conformal transformations \eqn{localconftrf} and \eqn{conftrfmatter}. When black hole Hawking radiation is
considered, different observers may have different ideas about what the vacuum state is, and as was argued in
Ref.~\cite{GtHcompl}, they may therefore also assign different values to the vacuum expectation value of the
field \(\w(x)\). Since the Ricci curvature \(\hat R_{\m\n}(x)\) is not at all invariant under the conformal
transformation \eqn{localconftrf}, these observers also disagree about the matter distribution in their
universe. It was argued in Ref.~\cite{GtHcompl} that this is inevitable: the outside observers will see the
need to include the back reaction of Hawking radiation to the r.h.s. of Einstein's field equation, while an
ingoing observer will not wish to include that. This is the central theme of the black hole complemantarity
issue. However, there now is one important complication: conformal anomalies. They should not be allowed to
ruin exact local conformal invariance. A local conformal transformation is necessary to transform from one
space-time to another where observers disagree about the ground state. In Appendix \ref{Weyl} it is also
derived how a local conformal transformation affects the stress-energy-momentum tensor of matter.
Appendix~\ref{conffequ} handles the field equations that would be generated by the Weyl-squared action. They
are found to be Einstein's equations with an energy momentum source that obeys equations by itself.

\newsecl{The divergent part of the \(\w\) integral in a curved background space-time}{scalefint}

Calculations related to the conformal term in gravity, and their associated anomalies, date back from the
early 1970s and have been reviewed amnong others in a nice paper by Duff\cite{Duff}. In particular, we here
focus on footnote (4) in that paper.\fn{This footnote reads: \({}^4\)If one starts with a classically
non-Weyl invariant theory (e.g. pure Einstein gravity) and artificially makes it Weyl invariant by
introducing via a change of variables \(g'_{\m\n}=e^{2\s(x)}g_{\m\n}(x)\) an unphysical scalar \emph{spurion}
\(\s(x)\), then unitarity guarantees that no anomalies can arise because this artificial Weyl invariance of
the quantum theory, \(g'_{\m\n}=\W^2(x)g'_{\m\n}(x)\) with \(e^{2\s(x)}\ra\W^2(x)e^{2\s(x)}\), is needed to
undo the field redefinition and remove the spurious degree of freedom. Professor Englert informed me in
Trieste that this is what the authors of Ref\cite{ETG} had in mind when they said that anomalies do not
arise. Let us all agree therefore that many of the apparent contradictions are due to this misunderstanding.
}

First, we go to \(n\) space-time dimensions, in order later to be able to perform dimensional
renormalization. For future convenience (see Eq.~\eqn{EHndim}), we choose to replace the parameter \(\w\)
then by \(\w^{2/(n-2)}\), so that Eq.~\eqn{gsplit} becomes
 \be g_{\m\n}(x)=\w^\fract 4{n-2}\,\hat g_{\m\n}(x)\ . \eel{gsplitndim}
Temporarily, we ignore the cosmological term, so that there is no \(\w^4\) interaction, in order to be able
to do perturbative calculations. In terms of \(\hat g_{\m\n}\) and \(\w\), the Einstein-Hilbert action
\eqn{EH}, with the matter contribution added, then reads
 \be S=\int\dd^nx\sqrt{-\hat g}\left({1\over 16\pi G_N}\bigg(\w^2\hat R+{4(n-1)\over n-2}\,\hat g^{\,\m\n}
 \pa_\m\w\,\pa_\n\w\bigg) +\LL^\mat(\hat g_{\m\n}),\w \right)\ , \eel{EHndim}
replacing Eq.~\eqn{EHsplit}.
We use the caret (\(\hat{}\)) to indicate all expressions defined by the tensor
\(\hat g_{\m\n}\), such as covariant derivatives, as if that were the true metric tensor.

Let us assume the absence of terms \emph{linear} in the matter fields \(\f\), which usually are removed
anyway by shifting the scalar fields. Then there are also no terms that are \emph{cubic} in the \(\w\) field.
The kinetic terms have been carefully constructed so as to be conformally invariant, hence
\(\w\)-independent. The mass terms are quadratic in \(\w\), and cubic scalar interaction terms, as well as
fermionic mass terms, are \emph{linear} in \(\w\). In this case, then, the \(\w\) integration is purely
Gaussian, and produces a determinant that can be computed precisely as a power series when the background
metric is curved.

Note that, in `Euclidean gravity', the \(\w\) integrand has the wrong sign. This is why \(\w\) must be chosen
to be on a complex contour, as we did in Eq.~\eqn{omegaeta}. In practice, it is easiest to do the functional
\(\w\) integration perturbatively, by writing
 \be \hat g_{\m\n}(x)=\eta_{\m\n}+\k\,h_{\m\n}(x)\ ,\qquad\eta_{\m\n}=\hbox{diag}(-1,1,1,1)\
 ,\qquad\k=\sqrt{8\pi G_N}\ , \eel{etaplush}
and expanding in powers of \(\k\) (although later we will see that that expansion can sometimes be summed).

We replace \(\w(x)\) by \(\eta(x)\) as in Eq.~\eqn{omegaeta}, where, in \(n\) dimensions, the constant
\(\tl\k\) is now given by

 \be\tl\k^2 ={2\pi G_N (n-2)\over n-1}\ . \eel{kappatilden}
 This turns the action \eqn{EHndim} into\fn{Note that, therefore, Newton's constant disappears nearly completely (its
use in Eq.~\eqn{etaplush} is inessential). It only returns when there are explicit non-conformal terms in the
matter lagrangian, such as mass terms. This a characteristic feature of this approach.}
 \be S=\int\dd^nx\sqrt{-\hat g}\left(-\half \hat g^{\m\n}\pa_\m\eta\pa_\n\eta-\half{n-2\over
 4(n-1)}\hat R\eta^2+\LL^\mat(\hat g_{\m\n},i\eta)\right)\ . \eel{omegaL}

We see that there is a kinetic term (perturbed by a possible non-trivial space-time dependence of \(\hat
g_{\m\n}\)), and a direct interaction, ``mass" term proportional to the background scalar curvature \(\hat
R\):
 \be {n-2\over 4(n-1)}\,\hat R\quad \buildrel{n\ra 4}\over{\longrightarrow}\quad \fract16 \hat R\ , \eel{Rterm}

The expressions for the one-loop diagrams of the \(\eta\) field diverge at \(n\ra 4\). By using general
covariance, one can deduce right away that the divergent terms must all combine in such a way that they only
depend on the Riemann curvature. Dimensional arguments then suffice to conclude that the coefficients must be
local expressions in the squares of the curvature.

The key calculations for the divergent parts have already been performed in 1973 \cite{GtHVeltman}. There, it
was found that a lagrangian of the form
 \be \LL=\sqrt{-g}\left(-\half g^{\m\n}(x)\,\pa_\m\vv\pa_\n\vv+\half M(x)\,\vv^2\right)\ , \eel{tHV}
will generate an effective action, whose divergent part is of the form \def\div{{\mathrm{\,div}}}
 \be  S^\div&=&\int\dd^nx\,\G^\div(x)\ ,\nn   \G^\div&=&{\sqrt{-g}\over
8\pi^2(4-n)}\left(\fract1{120}(R_{\m\n}R^{\m\n}-\fract13 R^2)+ \fract14
 (M+\fract16 R)^2\right)   \eel{scalarpole}
(we use here a slightly modified notation, implying, among others, a sign switch in the definition of the
Ricci curvature, and a minus sign as ref.~\cite{GtHVeltman} calculated the Lagrangian \(\LL+\D \LL,\
 \D \LL=-\G^\div\) needed to obtain a finite theory.)

In our case, we see that, in the Lagrangian \eqn{omegaL},
 \be M=-\fract16\hat R\ ,\qquad\G^\div={\sqrt{-\hat g}\over 960\pi^2(4-n)}(\hat R_{\m\n}\hat R^{\m\n}-\fract13 \hat R^2)\ ,
 \eel{divpart}
since the second term in \eqn{scalarpole} cancels out exactly. Indeed, it had to cancel out, as
Eq.~\eqn{divpart} has to reflect the conformal symmetry, see Eq.~\eqn{scaleinvaction} in Appendix~\ref{Weyl}.

To see what the divergence here means, we use the fact that the mass dependence of a divergent integral
typically takes the form
 \be  f(n)m^{n-4}\G(2-\half n)\ra {f(n)\over 4-n}\left(1+(n-4)\log\big({m\over\L}\big)\right)\ra
 f(4)\bigg(\log\L+{1\over 4-n}\bigg)&&\nn +\hbox{ finite}\  ,&& \eel{cutoff}
where \(m\) stands for a mass or an external momentum \(k\), and \(\L\) is some reference mass, such as an
ultraviolet cutoff. Thus, the divergent expression \(1/(4-n)\) generally plays the same role as the logarithm
of an ultraviolet cutoff \(\L\).

It is clear that an ultraviolet cut-off would violate local conformal invariance. Equivalently, one may note
that the theory is conformally invariant in 4 dimensions but not in \(n\) dimensions. A conformally invariant
counter lagrangian would then have to be of the form of Eq.~\eqn{WeylsqunA}:
  \be\D\LL={C(n)}\,\w^{2(n-4)\over n-2}\sqrt{-\hat g}\,\hat W_{\a\b\m\n}\hat W^{\a\b\m\n}\ ,
  \qquad C(n)={1\over 1920\,\pi^2\,(n-4)}\ .\eel{Weylsqun}
If we remove the \(n\) dependence in the power of \(\w\), conformal invariance is broken and an anomaly
emerges, an effective interaction proportional to \(\log(\w)\). If we keep the \(n\)-dependent power of
\(\w\) in Eq.~\eqn{Weylsqun} then our theory has an essential singularity (a branch cut) at \(\w=0\
(\eta=0)\). Field theories do not normally have such a singularity in the field dependence of the fundamental
interactions, and we may ask ourselves whether it is allowed here. Note that the region \(\w\ra 0\) describes
the zero distance limit of quantum gravity, a domain that is not well understood.

\newsecl{Divergences due to matter}{matterdiv} \def\conf{\mathrm{\,conf}}

The conformally invariant kinetic term for scalar fields \(\vv(x)\) is described by the action
 \be\LL^\vv_\conf=-\half\sqrt{-g}(g^{\m\n}\pa_\m\vv\,\pa_\n\vv+\fract16 R\vv^2)\ , \eel{confscalar}
where the second term is a by now familiar necessity for complete conformal invariance. The extra term with
the Ricci scalar is in fact the same as the term \eqn{Rterm} in Eq.~\eqn{omegaL}.

The contribution of these scalar fields to the divergences of the one-loop diagrams with only external \(\hat
g_{\m\n}\) lines is exactly the same as that of the \(\w\) (or \(\eta\)) fields, and the contribution of the
Maxwell or Yang Mills kinetic terms and the spinor kinetic terms in Eq.~\eqn{conformalL} can also be
calculated. The outcome of this calculation is well-known\cite{PvN1}---\cite{deser} and the calculation for
the spin \(0,\ \half\) and 1 case was once more recapitulated in \cite{GtHconfgrav}. If we have one \(\w\)
field, \(N_0\) real scalar field components, \(N_{1/2}\) elementary Majorana spinor fields (or \(\half
N_{1/2}\) Dirac fields), \(N_1\) real vector fields, \(N_{3/2}\) gravitinos and \(N_2\) spin 2 fields, the
total divergence is described by the effective action
 \be S^\eff=2C\int\dd^nx \sqrt{-\hat g}\bigg(\hat R^{\m\n}\hat R_{\m\n}-\fract13 \hat R^2\bigg)\ ,
 \eel{totaldiv} where
\be C={1\over 16\pi^2(4-n)}\left(\fract 1{120}(1+N_0)+\fract1{40}N_{1/2}+\fract 1{10}N_1 -\fract{233}{720}N_{3/2}
 +\fract{53}{45}N_2\right) \ . \eel{totalcoeff}
Here, the first 1 is the effect of the conformal component \(\w\) of gravity itself. In this paper we omit
the last two terms of this expression because they refer to not evidently renormalizable fields.\fn{They were
mentioned here just to note that there is a minus sign due to gravitinos, so that a resolution of the anomaly
problem might lie there, but it will not be further pursued.}

The contributions from all renormalizable fields clearly add up with the same sign. This, in fact, could have
been expected from simple unitarity arguments, but as such arguments famously failed when the one-loop beta
functions for different particle types were considered, it is preferred to do the calculation
explicitly.\cite{GtHconfgrav}

Our question now is what to do with this divergence. There are various options to be considered:
    \bi{$i$.} The action \eqn{EH} no longer properly describes the situation at scales close to the
Planck scale. At \(|k|\approx M_{Pl}\), we no longer integrate over \(\w(k)\), which has two consequences: a
natural cut-off at the Planck scale, and a breakdown of conformal invariance. Indeed, this would have given
the badly needed scale dependence to obtain a standard interpretation of the amplitudes computed this way, if
one would take the following viewpoint: consider the canonical action, which is the sum of \eqn{EHsplit} to
\eqn{Lint}, and consider the functional integral over the field \(\w\) that was just performed. Then one
observes that, in the effective action thus obtained, all reference to Newton's constant \(G_N\) is hidden in
the rescaled massive parameters of the renormalized matter lagrangian. One might welcome the idea that an
explicit breakdown of conformal invariance, in terms of anomalies, may produce more non-trivial structure in
the theory.

This option, however, we dismiss, because for the discussion of black holes, exact local conformal invariance
is needed explicitly. As in a Higgs theory, there is nothing wrong with attributing \emph{all} observed
breakdown of the symmetry (here conformal symmetry) to the Higgs mechanism.
 \itm{$ii$.} Have all divergences cancel out. In some supergravity theories, the conformal anomaly
indeed cancels out\cite{Fradkin} to zero.\fn{I thank M.~Duff for this observation.} The problem here is that
unitarity is questionable in these theories. As long as we limit ourselves to conventional, renormalizable,
matter fields, their contributions, together with the contribution of the \(\eta\) field, all add up with the
same sign. Of course, interaction effects were not yet included in our calculation, but it seems unlikely
that they could give any relief; in that case, one would definitely have to require the interactions to be
strong, a murky terrain of this theory in any case.

It should be added that, if the one loop diagrams for the spin 0, \(\half\) and 1 fields would all turn out
either to be finite or renormalizable in the standard way, the situation for the remaining fields \(\hat
g_{\m\n}\) would still be exotic. As there are no propagator terms at all in the classical lagrangian
\eqn{EHsplit} or \eqn{EHndim}, the propagator and the interaction terms would all have to come from the loop
diagrams. They would be conformally invariant but that would imply an effective kinetic term with four
derivatives, hence an effective \(1/k^4\) propagator that could easily develop unphysical, Landau-like poles.
Thus, with this option, we are not out of the woods.
 \ei Incidently, this feature shows that canonical quantum gravity has much in common with Higgs theories in
the Higgs phase, the \(\w\) field playing the role of the Higgs field with non-vanishing vacuum expectation
value, while renormalizability of the system is ruined by the absence of a decent kinetic term in the action
for the degrees of freedom in the symmetric mode. This suggests the following alternative scenario:
 \bi{$iii$.} Add a kinetic term for the \(\hat g_{\m\n}\) field in the action. This
must be locally conformally invariant, and therefore only the expression \eqn{totaldiv}, with a finite
coefficient \(C\), qualifies. The problem here is that, regardless whether we are in the Higgs mode or not,
this kinetic term is not of the standard canonical type, being quartic in the space and time derivatives, so
that unitarity of the evolution operator is far from guaranteed. Refs.~\cite{PM},\cite{BM} made some brave
attempts, but we note that local conformal invariance does not admit a unique definition of energy (as this
is not a conformal invariant), so we cannot even begin constructing a Fock space with only positive energy
physical particle states.
 \itm{$iv$.} A fourth option is to keep the coefficient \(C\) of the Weyl action \eqn{totaldiv} \emph{very large},
say 20 to 40 orders of magnitude (it is dimensionless). In this case the coefficient for the graviton
propagator becomes very small. In the Higgs mode, the full propagator \(P_{\m\n|\a\b}(k)\) is the inverse of
an expression of the form
 \be \G(k)\approx -(k^2+Ck^4)\ , \eel{twopointf}
which will show a \(k\) dependence going like
 \be P(k)\ \approx\ {1\over k^2+C k^4}\ \approx\ {1\over k^2}-{1\over k^2+m^2}\ ;\qquad m^2=1/C\ . \eel{cancelprop}
This means that, at distances smaller than \(\sqrt{C}\) in Planck units, the gravitational force is screened,
by opposite sign massive gravitons. On the one hand one might object that such wrong sign gravitons violate
unitarity, but this time the theory is nearly classical (all gravitational loop corrections are down by many
orders of magnitude compared to the tree diagrams, \emph{at all scales}, so that, at all scales, violation of
unitarity is minute. Even if this theory cannot be correct formally, one might argue that inconsistencies at
the \(20^\mathrm{th}\) decimal place or beyond might be something to worry about later.

Small distance screening of the gravitational force may be an interesting phenom\-enon to be
looked for experimentally.
 \itm{$v$.} Finally, more realistically perhaps, one might simply decide to leave the problem of the
functional integral over the \(\hat g_{\m\n}\) fields open for later. This author suspects that a full theory
will go beyond simple-minded quantum field theories, and even calls the validity of quantum mechanics itself
(or more precisely, its usual Copenhagen interpretation) into question here, but as for this paper, the issue
will not be further pursued. \ei
 It may or may not be that the local conformal anomaly can be removed. We do point out that the anomaly
discussed in this section only occurs when the background metric has curvature (either Weyl or Ricci), in
which case there is a big distinction between local conformal transformations and space-time
reparametrizations. In the next section, the anomaly in flat space-time is considered. There, we can do much
more, as we shall see.

\newsecl{The case of a flat background.}{flat}

Let us assume that the matter fields \(\f^\mat\) consist of Yang-Mills fields \(A_\m^a\), Dirac fields
\(\bar\j,\ \j\) and scalar fields \(\vv\), the latter three sets being in some (reducible or irreducible,
chiral or non chiral) representation of the local Yang-Mills gauge group. As in the previous sections,
\emph{gravitinos}, or spin \(\fract32\) fields, are not included in this investigation, conceivably a
mistake, perhaps to be remedied in future investigations.\fn{Note that supersymmetry is not a basic
ingredient of this theory; indeed, the conformal invariance of theories with supersymmetry appears to be an
obstacle rather than an asset in our approach, as we shall see.} In this section, we have a flat
``background" metric \(\hat g_{\m\n}\). Note however, that the original metric \(g_{\m\n}=\w^2(x)\,\hat
g_{\m\n}\), is \emph{not} flat.

For brevity, we will write complex scalar fields as pairs of real fields, and if Weyl or Majorana fermions
occur, the Dirac fields can be replaced by pairs of these.\fn{A single Weyl or Majorana fermion then counts
as half a Dirac field.} Let us rewrite the lagrangian for matter interacting with gravity more precisely than
in Sections~\ref{confmatter.sec} and \ref{matterdiv}:
 \be \LL(\hat g_{\m\n},\eta,\f^\mat)&=& -\quart G^a_{\m\n}G^a_{\m\n}-\bar\j\,\hat\g^\m \hat D_\m\,\j-\half \hat g^{\m\n}
 (D_\m\vv\,D_\n\vv+\pa_\m\eta\,\pa_\n\eta) \nn && -\fract1{12}\hat R(\vv^2+\eta^2)-V_4(\vv)-iV_3(\vv)\eta+\half\tl\k^2
 m_i^2\eta^2\vv_i^2-\tl\L\eta^4 \nn &&-\bar\j(y_i\vv_i+iy_i^5\g^5\vv_i+i\tl\k m_d\eta)\j\ , \eel{fullL}
where the \(\w\) field was replaced by the \(\eta\) field, with Eq.~\eqn{omegaeta}, \(G_{\m\n}\) is the (non
Abelian) Yang-Mills curvature, and \(D_\m\) and \(\hat D_\m\) are covariant derivatives containing the
Yang-Mills fields; \(\hat\g_\m\) and \(\hat D_\m\) also contain the vierbein fields and connection fields
associated to \(\hat g_{\m\n}\); the Yukawa couplings \(y_i\), \(y_i^5\) and fermion mass terms \(m_d\) are
matrices in terms of the fermion indices. The scalar self interactions, \(V_3(\vv)\) and \(V_4(\vv)\) must be
a third and fourth degree polynomials in the fields \(\vv_i\):
 \be V_4(\vv)&=&\fract1{4!}\l\vv^4\iss\fract1{4!}\l^{ijk\ell}\vv_i\vv_j\vv_k\vv_\ell\ ;\crl{fourfield}
 V_3(\vv)&=&\fract1{3!}\tl g_3^{ijk}\vv_i\vv_j\vv_k\ ,\qquad \tl g_3=\tl\k g_3\ . \eel{threefield1}
In Eq.~\eqn{fullL}, like \(m_d\), also \(m_i^2\d_{ij}\) are (not necessarily positive) mass matrices, in
general. Furthermore, \(\tl\L\) stands for \(\fract16\tl\k^2\L\). Of course, all terms in \eqn{fullL} must be
fully invariant under the Yang-Mills gauge rotations. They must also be free of Adler Bell Jackiw
anomalies.\cite{ABJ}

Now that the dilaton field \(\eta\) has been included, the entire lagrangian has been made conformally
invariant. It is so by construction, and no violations of conformal invariance should be expected. This
invites us to consider the beta functions of the theory. Can we conclude at this point that the beta
functions should all vanish?

Let us not be too hasty. In the standard canonical theory, matter fields and their ineractions are
renormalized. Let us consider dimensional renormalization, and the associated anomalous behavior under
scaling. In \(4-\e\) dimensions, where \(\e\) is infinitesimal, the scalar field dimensions are those of a
mass raised to the power \(1-\e/2\), so that the couplings \(\l\) have dimension \(\e\). This means that in
most of the terms in the lagrangian~\eqn{fullL} the integral powers of \(\eta\) will receive extra factors of
the form \(\eta^{\pm\e}\) or \(\eta^{\pm\e/2}\), which will then restore exact conformal invariance at all
values for \(\e\). If we follow standard procedures, we accept that \(\eta\) is close to \(-i/\tl\k\), so
that singularities at \(\eta\ra0\) or \(\eta\ra\infty\) are not considered to be of any significance. Indeed,
the limit \(\eta\ra0\) may be seen to be the small-distance limit. This is the limit where gravity goes wrong
anyway, so why bother?

However, now one could consider an extra condition on the theory. Let us assume that the causal structure,
that is, the location of the light cones, is determined by \(\hat g_{\m\n}\), and that there exist dynamical
laws for \(\hat g_{\m\n}\). This was seen to be a very useful starting point for a better understanding of
black hole complementarity\cite{{GtHcompl}}. The laws determining the scale \(\w(x)\) should be considered to
be dynamical laws, and the canonical theory of gravity itself would support this: formally the functional
integral over the \(\eta\) fields is exactly the same as that for other scalar fields.

In view of the above, we do think it is worthwhile to pursue the idea that the \(\eta\) field must be handled
just as any other scalar component of the matter fields; but then, in the \(\e\ra0\) limit, after
renormalization, fractional powers would lead to \(\log(\eta)\) terms, and these must clearly be excluded.
Renormalization must be done in such a way that no traces of logarithms are left behind. Certainly then, a
scale transformation, which should be identical to a transformation where the fields \(\eta\) are scaled,
should not be associated with anomalies. Implicitly, this also means that the region \(\eta\ra 0\) is now
assumed to be regular. This is the small distance region, so that, indeed, our theory says something
non-trivial about small distances. This is why our theory leads to new predictions that eventually should be
testable. Predictions follow from the demand that all beta functions of the conformal ``theory"~\eqn{fullL}
must vanish.

We emphasize that this does not mean that the matter lagrangian itself must have vanishing \(\b\) functions.
It may have mass terms and other dimensionfull couplings, which just means that the \(\eta\) field couples
nontrivially to matter. Only after adding the \(\eta\) field, we now demand that the \(\b\) functions all
vanish.

Note that one set of terms is absent in Eq.~\eqn{fullL}: the terms linear in \(\vv\) and hence cubic in
\(\eta\). This,
 of course, follows from the fact that, usually, no terms linear in the scalar fields are needed in the standard matter
 lagrangians; such terms can easily be removed by translations of the fields: \(\vv_i\ra\vv_i+a_i\) for some constants
 \(a_i\). Thus, the classical lagrangian is stationary when the fields vanish: \(\vv=0\) is a classical solution. In
our present notation, this observation is equivalent to the observation that fields may be freely transformed
into one another without modifying the physics. One such transformation is a rotation of one of the scalar
fields, say \(\vv_1\), into the \(\eta\) field:
  \be \Big\{\matrix{\vv_1 &\ra& \vv_1\,\cosh\a_1+i\eta\,\sinh\a_1 \ ,\cr \eta&\ra&\eta\,\cosh\a_1-i \vv_1\,\sinh\a_1\ ,}
  \eel{etarotate}
where \(\a_1\) stands for the original shift of the field \(\vv_1\). The transformation is taken to be a
hyperbolic rotation because the ``kinetic term" \(-\half(\pa\eta^2+\pa\vv^2)= \half(\pa\tl\w^2-\pa\vv^2)\) in
Eq.~\eqn{fullL} has to be invariant; \(\et\) is imaginary.

In the ``unitarity gauge" \(\eta=-i/\tl\k\), corresponding to \(\w=1\), we see that the scalar field
\(\vv_1\) is shifted by a constant. The second transformation, that of \(\eta\) in Eq.~\eqn{etarotate}, is a
simple redefinition of the \(\w\) field that can be made undone by a local conformal transformation, so it
has little physical significance; it is there because of the conformal coupling
\(-\fract1{12}R(\eta^2+\vv^2)\) in Eq.~\eqn{fullL}. In most cases, these transformations need not be
considered since terms linear in \(\vv\) will in general not be gauge invariant.

Notice also that the Yang-Mills fields are not directly coupled to the \(\eta\) field. If we stay close to
the ``unitarity gauge", \(\eta\ra -i/\tl\k\), we can see why this is so. Since not \(\eta\), but \(\tl\w\) is
real, the invariant quantity is \(\vv^2-\tl\k^2\w^2\). Rotating \(\vv\) fields with \(\eta\) fields would
therefore be a non-compact transformation, and Yang Mills theories with non-compact Lie groups usually do not
work. There is food for thought here, but as yet we shall not pursue that.

\newsecl{The \(\b\) functions}{beta}

Thus, we return to a theory to which all known quantum field theory procedures can be applied, the only new
thing being the presence of an extra, gauge neutral, spinless field \(\eta\), and the perfect local scale
invariance of the theory.

We arrived at the lagrangian~\eqn{fullL}, and we wish to impose on it the condition that all its beta
functions vanish, since local conformal invariance has to be kept. As the theory is renormalizable, the
number of beta functions is always exactly equal to the number of freely adjustable parameters. In other
words: we have exactly as many equations as there are freely adjustable unknown variables, so that \emph{all
coupling constants, all mass terms and also the cosmological constant,} should be completely fixed by the
equations \(\b_i=0\). They are at the stationary points. Masses come in the combination \(\tl\k\,m_i\) and
the cosmological constant in the combination \(\tl\k^2\L\), so all dimensionful parameters of the theory will
be fixed in terms of Planck units.

In principle, there is no reason to expect any of these fixed points to be very close yet not on any of the
axes, so neither masses nor the cosmological constant can be expected to be unnaturally small, at this stage
of the theory. In other words, as yet no resolution of the hierarchy problem is in sight: why are many of the
physical mass terms 40 orders of magnitude smaller than the Planck mass, and the cosmological constant more
than 120 orders of magnitude? We have no answer to that in this paper, but we shall show that the equations
are quite complex, and exotic solutions cannot be excluded.

Also, the existence of infinitely many solutions cannot be excluded. This is because one can still adjust the
composition and the rank(s) of the Yang-Mills gauge group(s), as well as one's choice of the scalar and
(chiral) spinor representations.\fn{which of course must be free of Adler Bell Jackiw
anomalies\cite{ABJ}\cite{bardeen}.} These form infinite, discrete sets. However, many choices turn out not to
have any non trivial, physically acceptable fixed point at all: the interaction potential terms \(V(\vv)\)
must be real and properly bounded, for instance. Searches for fixed points then automatically lead to
vanishing values of some or more of the coupling parameters, which would mean that the symmetries and
representations have not been chosen correctly.

Every advantage has its disadvantage. Since all parameters of the theory will be fixed, we cannot apply
perturbation theory. However, we can make judicious choices of the scalar and spinor representations in such
a way that the existence of a fixed point for the gauge coupling to these fields can be made virtually
certain. The \(\b\) function for \(SU(N)\) gauge theories with \(N_f\) fermions and \(N_s\) complex scalars
in the elementary representation is
 \be 16\pi^2\,\b(g)&=&-a\,g^3-(b_1\,g^5+b_2\,y^2g^3)+\OO(g^7)\ ,\crl{betathree}
 a&=&\fract {11}3 N-\fract23 N_f-\fract16 N_s\ , \crl{betag}
 b_i&=&\OO(N^2,\ N\,N_f,\ N\,N_s)\ . \eel{betatwoloop}
Choosing one scalar extra, or one missing, we can have \(a\) as small as \(a=\pm\fract16\), while a quick
inspection in the literature\cite{twoloops}\cite{threeloops} reveals that, in that case, \(b_i\) may still
have either sign:\fn{This rules out those supersymmetric approaches where the \(\b\) function would vanish at
all orders.}
 \be b_i=\pm\OO(N^2)\ . \eel{twoloopsign}
depending on further details, such as the ratio of fermions and scalars, the presence of other
representations, and the values of the Yukawa couplings. Choosing the sign of \(a\) opposite to that of
\(b\), one then expects that a fixed point can be found at\fn{The Yukawa coupling \(y\) will be \(\OO(g)\),
so that the \(b\) terms together can be handled as a \(b\,g^5\) term.}
 \be g^2=-a/b=\OO(1/N^2)\ . \eel{pertfixedpoint}

This implies that the relevant coupling at large \(N\), which is \(\tl g^2=g^2N\), can also be made small,
and hence it is reasonable to presume that the following procedure is reliable. Let there be \(\n\) physical
constants, the \(\n^\mathrm{\,th}\) one being the gauge coupling \(g\), which is determined by the above
equation \eqn{pertfixedpoint}. If we take all other coupling parameters to be of order \(g\) or \(g^2\), then
the remaining beta function equations are reliably given by the one-loop expressions only, which we shall
give below. Now these are \(\n-1\) equations for the \(\n-1\) remaining coupling parameters, and they are now
inhomogeneous equations, since the one coupling, \(g^2\), is already fixed. All we have to do now, is find
physically acceptable solutions. We already saw that non-Abelian Yang-Mills fields are mandatory; we shall
quickly discover that, besides the \(\eta\) fields, both fermions and other scalar matter fields are
indispensable to find any non-trivial solutions.

It is illustrative to mention a solution that might spring to mind: \(\NN=4\) super Yang-Mills. We take its
lagrangian, and add to that the \(\eta\) field while postulating that this \(\eta\) field does not couple to
the \(\NN=4\) matter fields at all. Then indeed all \(\b\) functions vanish\cite{susyym}. However, since the
\(\eta\) field is not allowed to couple, the physical masses are all strictly zero, which disqualifies the
theory physically. Note, however, that also the cosmological constant is rigorously zero. Perhaps the
procedure described above can be applied by modifying slightly the representations in this theory, so that a
solution with masses close to zero, and in particular a cosmological constant close to, but not exactly zero,
emerges. Approaches along such lines have not yet been investigated further.

The one loop \(\b\) functions are generated by an algebra\cite{GtHalgebra}, in which one simply has to plug
the Casimir operators of the Yang Mills Lie group, the types of the representations, the quartic scalar
couplings and the fermionic couplings. If we write the scalar fields as
 \be\s_0(x)=\eta(x)\ ,\qquad \s_i(x)=\vv_i(x)\ ,\quad i=1,\cdots,\ n_\vv\ , \eel{scalardef}
then the generic lagrangian can be written as
 \be\LL=-\quart G_{\m\n}^a G_{\m\n}^a -\half(D_\m\s_i)^2-V(\s)-\bar\j\big(\g D+(S_i+i\g^5P_i)\s_i\big)\j\ , \eel{genL}
where \(\s_i\) and \(\bar\j,\ \j\) are in general in reducible representations of the gauge group, \(D_\m\)
is the gauge covariant derivative, \(V(\s)\) is a gauge-invariant quartic scalar potential, and \(S_i\) and
\(P_i\) are matrices in terms of the fermion flavor indices. Everything must be gauge invariant and the
theory must be anomaly free\cite{ABJ}\cite{bardeen}.

The covariant derivatives contain the hermitean representation matrices \(T^a_{ij}\), \(U^{L\,a}_{\a\b}\) and
\(U^{R\,a}_{\a\b}\):
 \be D_\m\s_i&\equiv&\pa_\m\s_i+iT^a_{ij}A_\m^a\s_j\ ;\crl{scalarcovder}
  D_\m\j_\a&\equiv&\pa_\m\j_\a+i(U^{L\,a}_{\a\b}P^L+U^{R\,a}_{\a\b}P^R)A_\m^a\j_\b\ ;\quad P^{L,R}
  \equiv\fract12(1\pm\g^5)\ . \eel{spinorcovder}
The gauge coupling constant(s) \(g\) are assumed to be included in these matrices \(T\) and \(U\). The
operators \(P^L\) and \(P^R\) are projection operators for the left- and right handed chiral fermions.

The group structure constants \(f^{abc}\) are also assumed to include a factor \(g\), and they are defined by
 \be [T^a,\,T^b]=if^{abc}T^c\ ;\quad [U^{L\,a},\,U^{L\,b}]=if^{abc}U^{L\,c}
 \ ;\quad [U^{R\,a},\,U^{R\,b}]=if^{abc}U^{R\,c}\ .{\quad} \eel{structurecnsts}
Casimir operators \(C_g,\ C_s\) and \(C_f\) will be defined as
 \be f^{apq}f^{bpq}=C^{ab}_g\ ,\quad \Tr(T^aT^b)=C_s^{ab}\ ,\quad \Tr(U^{L\,a}U^{L\,b}+U^{R\,a}U^{R\,b})=C_f^{ab}\ .
 \eel{casimir}
All these algebraical numbers were defined such that they are either real or hermitean.

All beta functions are given by writing down how the entire lagrangian~\eqn{genL} runs as a function of the
scale \(\m\)\cite{GtHalgebra}:
 \be{\m\pa\over\pa\m}\LL&=&\b(\LL)\iss{1\over 8\pi^2}\D\LL\ , \crl{betaLdef} 
\D\LL&=& -\quart G_{\m\n}^aG_{\m\n}^b(\fract{11}3C_g^{ab}-\fract16C_s^{ab}-\fract23C_f^{ab})-\D V-\bar\j(\D
S_i+i\g^5\D P_i)\s_i\,\j\ . \eel{deltaL} Here,
 \be\D V&=&\quart V_{ij}^2-\fract32 V_i(T^2\s)_i+\fract34(\s T^aT^b\s)^2\ +\nn
 &&\s_iV_j\Tr(S_iS_j+P_iP_j)-\Tr(S^2+P^2)^2+\Tr[S,P]^2\ ,  \eel{deltaV}
 where \(\ V_i=\pa V(\s)/\pa\s^i\ ,\quad V_{ij}\iss\pa^2V(\s)/\pa\s_i\pa\s_j \)\ .

 It is convenient to define the complex matrices \(W_i\) as
 \be W_i=S_i+iP_i\ ,\qquad \tl W_i=S_i-iP_i  \eel{Wmatrices}
(Note that, with the field \(\s_0=\et\), which has imaginary Yukawa couplings, the functions \(\tl W_i\) and
\(W_i^\dagger\) need not be the same). Then,
 \be \D W_i&=&\D S_i+i\D P_i\iss \quart W_k\tl W_k W_i+\quart W_i\tl W_k W_k+W_k\tl W_i W_k\  \nn
 &&\qquad -\ \fract32 (U^R)^2W_i-\fract32 W_i(U^L)^2+W_k\Tr(S_kS_i+P_kP_i)\ . \eel{deltaW}

If now we write the collection of scalars as \(\{\s_i= \vv_i,\ \s_0=\eta\}\), taking due notice of the
factors \(i\) in all terms odd in \(\eta\), we can apply this algebra to compute all \(\b\) functions of the
lagrangian~\eqn{fullL}.

The values of the various \(\b\) functions depend strongly on the choice of the gauge group, the
representations, the scalar potential function and the algebra for the Yukawa terms, and there are very many
possible choices to make. However, the signs of most terms are fixed by the
algebra~\eqn{deltaL}---\eqn{deltaW}. By observing these signs, we can determine which are the most essential
algebraic constraints they impose on possible solutions. As we shall see, they are fairly restrictive. They
are severely restrictive if one demands that the cosmological constant be excessively
small.\cite{GtHconfgrav2}

\newsecl{Solving the equations \(\b=0\).}{betanul}

Consider the dilaton field \(\eta\) added to the lagrangian, as defined in Eq.~\eqn{scalardef}. This requires
extending the indices \(i,\,j,\,\dots\) in the lagrangian \eqn{genL} to include a value \(0\) referring to
the \(\eta\) field. The unusual thing is now that the terms odd in \(\eta\) are purely imaginary, while all
terms in Eq.~\eqn{fullL} are of dimension 4. The effect of this anomalous sign in the beta function equations
is a slight complication. It illustrative to consider the general question: can one expect interesting
solutions to such equations and how can they be searched for systematically? Here, we briefly outline a
general approach that is just slightly different from the one in Ref~\cite{GtHconfgrav2}.

First, we follow the procedure outlined in Sect.~\ref{beta}, to define the Yang-Mills self coupling(s) \(g\).
We assume that a fixed point is found where \(g\) is fairly small, so that the remaining equations can be
handled at the one-loop level, to receive small higher loop corrections at a later stage. Of course, theories
with larger couplings will be accordingly more difficult to handle. After our program to construct the values
of the other couplings (which all will be proportional to \(g\) or \(g^2\)), we will have to check whether
\(a\) and \(b\) in Eqs.~\eqn{betag} and \eqn{betatwoloop} indeed have opposite signs, giving a reasonably
small value for \(g\).

We then observe that the Yukawa couplings (where the \(\eta\) field controls the Dirac, or Majorana, masses), obey
equations only containing the other Yukawa couplings and the gauge constant(s) \(g\), according to Eq.~\eqn{deltaW}. In
that equation, the terms containing \((U^R)^2\) and \((U^L)^2\) are linear in \(W\) and the rest are cubic in \(W\).
Quite generally, these equations are of the form
 \be Y^3-g^2\,Y=0\ . \eel{yakawafix}
In particular, the signs in these equations appear to be favorable to the existence of solutions.

It is helpful that an extremum principle exists. We can find a scalar function \(H\) of the Yukawa couplings \(W_i,\
\tl W_i\) of the form
 \be &H\iss \Tr\Bigg(\a (W_i \tl W_i)^2+\a(\tl W_iW_i)^2 +\b W_i\tl W_kW_i\tl W_k\nn&-\g\bigg((U^R)^2W_i\tl W_i+
  (U^L)^2\tl W_iW_i\bigg)+  \d \Tr(W_i\tl W_j+W_j\tl W_i)^2 \Bigg)\ ,& \eel{Hfunct}
such that, if an infinitesimal variation \((\d W_i \ ,\ \d W_i^*) \) is chosen, the variation \(\d H\) of
\(H\) is given by
 \be\d H=\e\Tr(\d W_i^*\D W_i+\D W_i^*\d W_i)\ , \eel{deltaH} with\(\D W_i\)
as given by Eq.~\eqn{deltaW}. One finds that the coefficients \(\a\)---\(\d\) have to be chosen as follows:
 \be \a=\fract18\ ,\quad\b=\half\ ,\ \g=\fract32\ ,\ \d=\fract18\ . \eel{coeff}
Thus, the problem of finding a fixed point here is reduced to the search of an extremum of \(H\). It is easy to see
that \(H\) must have a minimum if all Yukawa couplings are real, so that \(\tl W_i=W_i^*\). This minimum is away from
zero because at small values of \(W,\ \tl W\), the dominant quadratic term is negative. However, the contribution of
the \(\et\) field can easily be included in the argument:

First of all, with the \(\et\) field included, the function \(H\) is still real, because the terms \(W_0\)
and \(\tl W_0\) only occur in terms with an even number of \(\eta\) fields.

\noindent Secondly, the terms quartic in \(W_0\) and \(\tl W_0\) are still positive, because \(i^4=+1\).

\noindent Next, the coefficients in \(W_0,\ \tl W_0\), do contribute to terms quadratic in these coefficients
that have the wrong sign, if there are non-chiral fermions. This implies that, for large values of \(W_0\),
we have no proof yet that solutions exist. There might be an absolute minimum of \(H\) at \(W_0=0\). Perhaps
this leaves an interesting intermediate case with \(W_0\) extremely small, which may give perspectives for
approaches towards the hierarchy problem.

Finally then, we have to contemplate the scalar interactions. In Eq.~\eqn{deltaV}, the gauge and Yukawa couplings are
now fixed by the previous derivations. It remains to determine whether or not a quartic function \(V(\s)\) exists that
obeys the equation \(\D V=0\). There are as many equations as there are unknowns.

Let us temporarily ignore the dependence on the \(\et\) field and write Eq.~\eqn{deltaV} as
 \be \D V=\D_1-\D_2+\D_3+\D_4-\D_5-\D_6\ , \eel{deltaVi}
where the signs of the various terms are explicitly indicated (the sign in front of \(\D_6\) is negative
because the commutator there is antihermitean). Inspecting the equation a bit more closely, we see that \(\D
V\) would become positive for large values of \(V\), since then the first term, \(\D_1\), dominates ---
although we have the same problems as before concerning the \(\eta\) dependent terms. For small values of
\(V\) the terms \(\D_3-\D_5-\D_6\) dominate. A solution is likely to exist if these add up to be negative,
which means that the Yukawa couplings must dominate over the gauge couplings in \(\D_3\). Including now the
\(\et\) field complicates matters, but if their contributions are not too big, it seems likely that solutions
to the equation \(\D V=0\) can be found in many cases. Thus, all coupling constants, mass terms and
cosmological term will then be uniquely, or nearly uniquely fixed, in the sense that there will be theories
with one solution, theories with two or more solutions, but also many theories with no fixed point at all.

The values obtained for the physical parameters are quite general solutions of the above non-linear
equations, and there seems to be no reason why any of these parameters should be excessively small compared
to some others. We did investigate in which case the cosmological constant will be zero or excessively small.
One easily derives\cite{GtHconfgrav2} the equation
 \be 36\tl\L^2= 4\tl\L\,\Tr(m_d^2)+\Tr(m_d^4)- \quart\sum_im_i^4 \ . \eel{cosmocancel}
The last two terms of this equation resemble a supertrace. Since the sign of the l.h.s. must be positive, we read off
right away that there \emph{must} be fermions. If furthermore we like to have a very small or vanishing cosmological
constant \(\L\), we clearly need that the sum of the fourth power of the Dirac masses (approximately) equals the sum of
the fourth powers of the masses of the real scalar particles divided by 4.

If there were no scalar fields \(\vv_i\), Eq.~\eqn{cosmocancel} would have a solution, but, since the last
term would then be absent, the cosmological constant would come out fairly large.

It appears that in today's particle models not only the cosmological constant \(\tl\L\) but also the mass
terms are quite small, in the units chosen, which are our modified Planck units. Also, if there is a triple
scalar coupling, \(V_3(\vv)\), it appears to be small as well. This is the hierarchy problem, for which we
cannot offer any solution other than suggesting that we may have to choose a very complex group structure ---
as in the landscape scenarios often proposed in superstring theories. Perhaps, the small numbers in our
present theory are all related.

If the masses are indeed all small, then the only large terms in our equations are the ones that say how the
coupling constants and masses run with scale. Our theory suggests that they might stop running at some scale;
in any case, a light Higgs particle indeed follows from the demand that the Higgs self coupling is near an UV
fixed point.

Searches for more explicit models, including ones with small masses and cosmological constant, will be left
for the future.

\newsecl{Discussion.}{disc}

Our theory derives constraints from the fact that matter fields interact with gravity. The basic assumption
could be called a new version of relativity: the scalar matter fields should not be fundamentally different
from the dilaton field \(\eta(\vec x,t)\). Since there are no singular interactions when a scalar field tends
to zero, there is no reason to expect any singularity when \(\eta(\vec x,t)\) tends to zero at some point in
space-time. Standard gravity theory does have singularities there: this domain refers to the short distance
behavior of gravity, which is usually considered to be ``not understood". What if the short distance behavior
of gravity and matter fields is determined by simply demanding the absence of a singularity? Matter and
dilaton then join smoothly together in a perfectly conformally invariant theory. This, however, only works if
all \(\b\) functions of this theory vanish: its coupling parameters must be at a fixed point. There are only
discrete sets of such fixed points. Some theories have no fixed point at all in the domain where physical
constants are real and positive --- that is, stable. Searching for non trivial fixed points will be an
interesting and important exercise.

Our condition that local conformal invariance may not be broken explicitly but only spontaneously implies
that \emph{all} physical parameters, including the cosmological constant, will be fixed and calculable in
terms of the Planck units. This may be a blessing and a curse at the same time. It is a blessing because this
removes all dimensionless freely adjustable real numbers from our theory; everything is calculable, using
techniques known today; there is a strictly discrete set of models, where the only freedom we have is the
choice of gauge groups and representations. It is difficult to tell how many solutions there are; the number
is probably denumerably infinite.

This result is also a curse, because the values these numbers have in the real world is a strange mix indeed:
the range of the absolute values cover some 122 orders of magnitude:
 \be \tl\L=\OO(10^{-122})\ ;\qquad \m^2_\mathrm{Higgs}\approx 3.10^{-36}\ . \eel{numvalues}
The question where these various hierarchies of very large, or small, numbers come from is a great mystery
called the ``hierarchy problem". In our theory these hierarchies will be difficult to explain, but we do
emphasize that the equations are highly complex, and possibly theories with large gauge groups and
representations have the potential to generate such numbers.

Our theory is a ``top-down" theory, meaning that it explains masses and couplings at or near the Planck
domain. It will be difficult to formulate any firm predictions about physics at energies as low as the TeV
domain. Perhaps we should expect large regions on a logarithmic scale with an apparently unnatural scaling
behavior. There is in principle no supersymmetry, although the mathematics of supersymmetry will be very
helpful for constructing the first non-tivial models.

What is missing furthermore is an acceptable description of the dynamics of the remaining parts \(\hat
g_{\m\n}\) of the metric field. In Ref.~\cite{GtHconfgrav}, it was suggested that this dynamics may be non
quantum mechanical, although this does raise the question how \(\hat g_{\m\n}\) can back react on the
presence of quantum matter. Standard quantum mechanics possibly does not apply to \(\hat g_{\m\n}\) because
the notion of energy is absent in a conformal theory, and consequently the use of a hamiltonian may become
problematic. A hamiltonian can only be defined after coordinates and conformal factor have been chosen, while
this is something one might prefer not to do. The author believes that quantum mechanics itself may have to
be carefully reformulated before we can really address this problem.

Instead of keeping the coefficient in front of the conformal Weyl action infinite, we could also settle for a
value that is very large, say \(10^{40}\), but not infinite. This would reduce to zero all gravitational
effects at mass and energy scales beyond \(10^{-20}\) times the Planck mass, without modifying gravity as it
is known today, while all unitarity violations would stay well below the \(10^{-40}\) level. This would imply
that gravity is screened at short distances, with wrong sign, low mass ``anti"-gravitons; we then have a
theory that is only approximately unitary. All violations of unitarity would only show up at the
\(40^\mathrm{th}\) decimal place, even at the Planck scale; in a pragmatic world view one might be able to
live with that.

Our theory indeed is complex. We found that the presence of non-Abelian Yang-Mills fields, scalar fields and
spinor fields is required, while \(U(1)\) gauge fields are forbidden (at least at weak coupling, since the
\(\b\) function for the charges here is known to be positive). Because of this, one ``prediction" stands out:
there will be magnetic monopoles, although presumably their masses will be of the order of the Planck mass.

Finally, there is one other firm prediction: the constants of nature will indeed be truly constant. Attempts
to experimentally observe variations in constants such as the finestructure constant or the proton electron
mass ratio, with time, or position in distant galaxies, are predicted to yield negative results.

\newsecl{Conclusions}{concl} During the last decades, microscopic black holes were studied because they appeared to
cause conflicts in the conventional theory. Now, we tentatively conclude that this study may finally be paying off. Our
research was inspired by ideas launched recently\cite{GtHcompl}, where it was concluded that an effective theory of
gravity should exist in which the dilaton component either does not exist at all or is integrated out. This would
enable us to understand the black hole complementarity principle, and indeed, make black holes effectively
indistinguishable from ordinary matter at tiny scales. A big advantage of such constructions would be that, due to the
formal absence of black holes, we would be allowed to limit ourselves to topologically trivial, continuous spacetimes
for a meaningful and accurate, nonperturbative description of all interactions. This actually leads to a novel
constraint on theories, so that we get predictions affecting the Standard Model itself: constants of nature are truly
constant and they are, in principle, calculable, although the latter does require that we know exactly all fields and
their symmetry algebras. These are denumerable, and they can be guessed, perhaps.

Let us briefly summarize here how the present formulation can be used to resolve the issue of an apparent
clash between unitarity and locality in an evaporating black hole. An observer going into the hole does not
explicitly observe the Hawking particles going out. (S)he passes the event horizon at Schwarzschild time
\(t\ra\infty\), and from his/her point of view, the black hole at that time is still there. For the external
observer, however, the black hole has disappeared at \(t\ra\infty\). Due to the back reaction of the Hawking
particles, energy (and possibly charge and angular momentum) has been drained out of the hole. Thus, the two
observers appear to disagree about the total stress-energy-momentum tensor carried by the Hawking radiation.
Now this stress-energy-momentum tensor was constructed in such a way that it had to be covariant under
coordinate transformations, but this covariance only applies to \emph{changes} made in the
stress-energy-momentum when creation- and/or annihilation operators act on it. About these covariant changes,
the two observers do not disagree. It is the \emph{background subtraction} that is different, because the two
observers do not agree about the vacuum state. This shift in the background's source of gravity can be neatly
accommodated for by a change in the conformal factor \(\w(x)\) in the metric seen by the two observers.

This we see particularly clearly in Rindler space. Here, we can generate a modification of the background
stress-energy-momentum by postulating an infinitesimal shift of the parameter \(\l(x)\) in
Eqs.~\eqn{Ricciscaletrf} and \eqn{Ssol}. It implies a shift in the Einstein tensor \(G_{\m\n}\) (and thus in
the tensor \(T_{\m\n}\)) of the form \eqn{Gtensorshift} (see appendix \ref{Weyl}).

If now the transformation \(\l\) is chosen to depend only on the lightcone coordinate \(x^-\), then
 \be G_{--}\ra G_{--}-\pa_-^2\l\ , \eel{Gminminshift}
while the other components do not shift. Thus we see how a modification only in the energy and momentum of
the vacuum in the \(x^+\) direction (obtained by integrating \(G_{--}\) over \(x^-\)) is realized by a scale
modification \(\l(x^-)\).

In a black hole, we choose to modify the pure Schwarzschild metric, as experienced by an ingoing observer, by
multiplying the entire metric with a function \(\w^2(t)\) that decreases very slowly from 1 to 0 as
Schwarzschild time \(t\) runs to infinity. This then gives the metric of a gradually shrinking black hole as
seen by the distant observer. Where \(\w\) has a non vanishing time derivative, this metric generates a non
vanishing Einstein tensor, hence a non vanishing background stress-energy-momentum. This is the
stress-energy-momentum of the Hawking particles.

Calculating this stress-energy-momentum yields an apparently disturbing surprise: it does not vanish at apacelike
infinity. The reason for this has not yet completely been worked out, but presumably lies in the fact that the two
observers not only disagree about the particles emerging from the black hole, but also about the particles entering the
black hole, and indeed an infinite cloud of thermal radiation filling the entire universe around it.

All of this is a sufficient reason to suspect that the conformal (dilaton) factor \(\w(x)\) must be declared
to be locally unobservable. It is fixed only if we know the global spacetime and after choosing our
coordinate frame, with its associated vacuum state. If we would not specify that state, we would not have a
specified \(\w\). In `ordinary' physics, quantum fields are usually described in a flat background. Then the
choice for \(\w\) is unique. Curiously, it immediately fixes for us the sizes, masses and lifetimes of all
elementary particles. This may sound mysterious, until we realize that sizes and lifetimes are measured by
using light rays, and then it is always assumed that these light rays move in a flat background. When this
background is not flat, because \(\hat g_{\m\n}\) is non-trivial, then sizes and time stretches become
ambiguous. We now believe that this ambiguity is a very deep and fundamental one in physics.

Although this could in principle lead to a beautiful theory, we do hit a real obstacle, which is, of course,
that gravity is not renormalizable. This `disease' still plagues our present approach, unless we turn to
rather drastic assumptions. The usual idea that one should just add renormalization counter terms wherever
needed, is found to be objectionable. So, we turn to ideas related to the `primitive quantization' proposal
of Ref.~\cite{GtHquant}. Indeed, this quantization procedure assumes a basically classical set of equations
of motion as a starting point, so the idea would fit beautifully.

We were led to the class of models defined by the condition that local conformal symmetry is an exact
symmetry that is spontaneously broken. Therefore, in the symmetric phase, the \(\b\) functions must all
vanish. Valuable examples can be constructed as follows.
 \bi{Step 1.} A gauge field algebra and the associated fermionic and scalar representations are chosen in such a
way that the coefficient(s) \(a\) in Eq.~\eqn{betag} is/are small. We don't know the sign yet, so try both
possibilities.
 \itm{Step 2.} Write the Yukawa terms as matrices \(W_i,\ \tl W_i\) that, at lowest order, are proportional to the gauge
coupling constant(s) \(g\). Find these coefficients by extremizing \(H\) in Eq.~\eqn{Hfunct}.
 \itm{Step 3.} This enables us to compute the coefficients \(b_i\) in Eq.~\eqn{betatwoloop}. Check whether
\(a/b\) is negative and small. If not, try another algebra, going back to Step 1.
 \itm{Step 4.} Find a solution of the equation \(\D V=0\), Eq.~\eqn{deltaV}, \eqn{deltaVi}.
 \itm{Step 5.} Check the resulting masses and the cosmological constant. In physically realistic models
these should be very small. At this stage of our understanding, there seems to be no reason for these to be
small, so we expect only to achieve toy models, but hopefully more can be learnt by studying many
examples.\ei

Of course, many other questions are left unanswered. This is also why we call our result ``a class of
models". It is obviously important to have models without freely adjustable parameters. Our models are to be
constructed using the guideline that difficulties with exact local conformal invariance have to be addressed.
However, quite conceivably, further research might turn up more alternative options for a cure to these
difficulties. One of these, of course, is superstring theory. Superstring theory often leads one to avoid
certain questions to be asked at all, but eventually the black hole complementarity principle will have to be
considered, just as the question of the structure of Nature's degrees of freedom at distance and energy
scales beyond the Planck scale.

\section*{Acknowledgements}
The author thanks C.~Bachas, R.~Bousso, M.~Duff, G.~Dvali, S.~Giddings, C.~Kounnas, P.~Mannheim, C.~Taubes, and P.~van
Nieuwenhuizen for discussions.

\appendix
\newsecl{Local scale invariance and the Weyl curvature}{Weyl}

Consider an effective theory with not only general covariance,
 \be \hat g_{\m\n}\ra \hat g_{\m\n}+\hat D_\m u_\n+\hat D_\n u_\m\ , \eel{infcoordtrf}
where \(u_\m(x)\) are the generators of infinitesimal coordinate transformations, and \(\hat D_\m\) is the
covariant derivative with respect to \(\hat g_{\m\n}\); but now we also have a new kind of gauge invariance,
being local scale invariance, which we write in infinitesimal notation, for convenience:
 \be \hat g_{\m\n}\ra \hat g_{\m\n}+\l(x) \hat g_{\m\n}\ , \eel{localscale}
and we demand invariance under that as well. Note that this transformation is quite distinct from scale transformations
in the coordinate frame, which of course belongs to the coordinate transformations \eqn{infcoordtrf} and as such is
always an invariance of the usual theory. In short, we now have a theory with a 5 dimensional local gauge group.
Theories of this sort have been studied in detail\cite{PM}\cite{BM}.

The Riemann tensor \(\hat R^\a_{\ \b\m\n}\) transforms as a decent tensor under the coordinate transformations
\eqn{infcoordtrf}, but it is not invariant (or even covariant) under the local conformal transformation
\eqn{localscale}. Now, in four space time dimensions, we can split up the 20 independent components of the Riemann
tensor into the 10 component Ricci tensor
 \be \hat R_{\m\n}=\hat R^\a_{\ \m\a\n}\ , \eel{Ricci}
and the components orthogonal to that, called the Weyl tensor,
 \be   &&\hat W_{\m\n\a\b}=\hat R_{\m\n\a\b}+\nn &&{ }\hskip -30pt \half (-g_{\m\a}\hat R_{\n\b}+g_{\m\b}\hat
R_{\n\a}+g_{\n\a}\hat R_{\m\b}-g_{\n\b}\hat R_{\m\a})+\fract
 16 (g_{\m\a}g_{\n\b}-g_{\n\a}g_{\m\b})\hat R\ , \eel{Weyldef}
which is constructed in such a way that all its traces vanish, \(\hat W^\n_{\ \a\m\b}=0\), and therefore has
the just the remaining 10 independent components.

The transformation rules under coordinate transformations \eqn{infcoordtrf} are as usual; all these curvature
fields transform as tensors. To see how they transform under \eqn{localscale}, first note how the connection
fields transform:
 \be \hat \G_{\a\m\n}\ra(1+\l)\hat \G_{\a\m\n}+\half(\hat g_{\a\n}\pa_\m\l +\hat g_{\a\m}\pa_\n\l-\hat g_{\m\n}\pa_\a\l) + \OO(\l^2)\ , \eel{connectionscaletrf}
from which one derives
 \be \hat R_{\a\b\m\n}\ra (1+\l)\hat R_{\a\b\m\n}+\half(\hat g_{\a\n}\hat D_\b\pa_\m\l-\hat g_{\a\m}\hat D_\b\pa_\n\l-\hat
 g_{\b\n}\hat D_\a\pa_\m\l+\hat g_{\b\m} \hat D_\a\pa_\n\l)\ .
 \eel{Riemscaletrf}
From this we find how the Ricci tensor transforms:
 \be \hat R_{\m\n}\ra \hat R_{\m\n}-\hat D_\m\pa_\n\l-\half \hat g_{\m\n}\hat D^2\l\ ,\qquad \hat R\ra \hat
R(1-\l)-3\hat D^2\l\ . \eel{Ricciscaletrf} The Einstein tensor \(G_{\m\n}\) transforms as
 \be G_{\m\n}\ra G_{\m\n}- D_\m\pa_\n\l+g_{\m\n}D^2\l\ , \eel{Gtensorshift} which implies that the
 \emph{matter content} of space-time is not invariant under local conformal transformations.

The Weyl tensor \eqn{Weyldef}, being the traceless part, is easily found to be invariant (apart from the
canonical term):
 \be \hat W_{\a\b\m\n}\ra(1+\l)\hat W_{\a\b\m\n}\ . \eel{Weylscaletrf}
Since the inverse, \(\hat g^{\m\n}\), and the determinant, \(\hat g\), of the metric transform as
 \be \hat g^{\m\n}\ra(1-\l)\hat g^{\m\n}\ ;\qquad\hat g\ra (1+4\l)\hat g\ , \eel{detscaletrf}
we establish that exactly the Weyl tensor squared yields an action that is totally invariant under local
scale transformations in four space-time dimensions (remember that \(\hat g^{\m\n}\) is used to connect the
indices):
 \be\mathcal{L}=C\sqrt{-\hat g}\,\hat W_{\a\b\m\n}\hat W^{\a\b\m\n}=C\sqrt{-\hat g}(\hat R_{\a\b\m\n}\hat R^{\a\b\m\n}-
2\hat R_{\m\n}\hat R^{\m\n}+\fract 13 \hat R^2)\ ,
 \eel{Weylsqu}
which, due to the fact that the integral of
 \be \hat R_{\a\b\m\n}\hat R^{\a\b\m\n}-4\hat R_{\m\n}\hat R^{\m\n}+\hat R^2\eel{GB}
is a topological invariant, can be further reduced to
 \be\mathcal{L}=2C\sqrt{-\hat g}(\hat R_{\m\n}^2-\fract13 \hat R^2)\ , \eel{scaleinvaction}
to serve as a possible locally scale invariant Lagrangian.

The constant \(C\) may be any dimensionless parameter. Note that, according to Eq.~\eqn{Ricciscaletrf},
neither the Ricci tensor nor the Ricci scalar are invariant; therefore, they are locally unobservable at this
stage of the theory. Clearly, in view of Einstein's equation, \emph{matter}, and in particular its
stress-energy-momentum tensor, are locally unobservable in the same sense. This will have to be remedied at a
later stage, where we must work on redefining what matter is at scales much larger than the Planck scale.

If we replace the number of space-time dimensions from 4 to \(n\), we take
 \be g_{\m\n}=\w^{4\over n-2}\hat g_{\m\n}\ , \eel{metricsplitn}
and the expressions for the Weyl curvature become
  \be &&R_{\m\n\a\b}=W_{\m\n\a\b}+\nn &&{ }\hskip -30pt \fract
 1{n-2}(g_{\m\a}R_{\n\b}-g_{\m\b}R_{\n\a}-g_{\n\a}R_{\m\b}+g_{\n\b}R_{\m\a})+\fract
 1{(n-2)(n-1)}(-g_{\m\a}g_{\n\b}+g_{\n\a}g_{\m\b})R\ ; \eel{Weylndim}
the conformal invariant effective lagrangian then reads
  \be\LL=C\w^{2(n-4)\over n-2}\sqrt{-\hat g}\,\hat W_{\a\b\m\n}\hat W^{\a\b\m\n}\ . \eel{WeylsqunA}

\newsecl{conformal field equations}{conffequ}

\def\ns{\!\!} Here, we consider the field equations associated to the ``lagrangian" \eqn{scaleinvaction}. Consider an infinitesimal variation \(h_{\m\n}\) on the metric: \(\hat g_{\m\n}\ra \hat
g_{\m\n}+\d\hat g_{\m\n}\), \(\d\hat g_{\m\n}=h_{\m\n}\). The infinitesimal changes of the Ricci tensor and
scalar are
 \be \d \hat R_{\m\n}&=&\half(\hat D_\a \hat D_\m h^\a_\n+\hat D_\a \hat D_\n h^\a_\m -D^2 h_{\m\n}-\hat D_\m \pa_\n h^\a_\a)\
 ;\crl{Riccivary} \d \hat R&=&-h^{\a\b}\hat R_{\a\b}+\hat D_\a \hat D_\b h^{\a\b}-\hat D^2 h^\a_\a\ . \eel{scalarvary}
Using the Bianchi identity
 \be  D_\m  R^\m_{\ \n}=\half \pa_\n  R\ , \eel{Bianchi}
the variation of the Weyl action \eqn{Weylsqu}, \eqn{scaleinvaction} is then found to be
 \be &\d S= -2C\int\dd^nx \sqrt{-\hat g}\,h^{\a\b}\Box_{\,\a\b}^R\ ,\qquad\hbox{with}&\nn
 &\Box_{\,\a\b}^R=\hat D^2\hat R_{\a\b}-\fract13 \hat D_\a \hat D_\b \hat R-\fract16 g_{\a\b}\hat D^2 \hat R-2\hat
R^\m_\a \hat R_{\m\b}+2\hat R^{\m\n} \hat R_{\a\m\b\n}-\fract23 \hat R\hat R_{\a\b}\ . &\eel{Weylvary}

The classical equations of motion for the Ricci tensor as they follow from the Weyl action are therefore:
 \be \Box_{\,\a\b}^R=0\ . \eel{Weylequ}
To see their most salient features, let us linearize in \(\hat R_{\m\n}\) and ignore connection terms. We get
  \be \hat R_{\m\n}-\fract16\hat R \d_{\m\n}\deff S_{\m\n}\ ; \qquad \pa_\m S_{\m\n}=\pa_\n S_{\a\a}\
;\quad\pa^2 S_{\m\n}-\pa_\m\pa_\n S_{\a\a}=0\ . \eel{eomS}
 Defining \(\l(x)\) by the equation
 \be \pa^2\l\deff -S_{\a\a}\ , \eel{Lambdadef} we find that the solution \(S_{\m\n}\) of Eq.~\eqn{eomS} can be
written as
 \be S_{\m\n}=-\pa_\m\pa_\n\l+A_{\m\n}\ ,\quad\hbox{with}\quad\pa^2 A_{\m\n}=0\ ,\ A_{\a\a}=0\ ,\ \pa_\m
 A_{\m\n}=0\ . \eel{Ssol}
From Eq.~\eqn{Ricciscaletrf} we notice that the free function \(\l(x)\) corresponds to the local scale degree
of freedom \eqn{localscale}, while the equation for the remainder, \(A_{\m\n}\), tells us that the Einstein
tensor, after the scale transformation \(\l(x)\), can always be made to obey the d'Alembert equation \(\pa^2
G_{\m\n}=0\), which is basically the field equation for the stress-energy-momentum tensor that corresponds to
massless particles\fn{Not quite, of course. The statement only holds when these particles form classical
superpositions of plane waves such as an arbitrary function of \(x-t\).}. Thus, it is not true that the Weyl
action gives equations that are equivalent to Einstein's equations, but rather that they lead to Einstein
equations with only massless matter as their source.

\end{document}